\documentclass[preprint,authoryear]{elsarticle}

\journal{arXiv}

\usepackage{paralist}

\begin{document}

\begin{frontmatter}

\title{The \texttt{TimeMachine} for Inference on Stochastic Trees}

\author[gc]{Gianluca Campanella}
\ead{g.campanella11@imperial.ac.uk}
\author[mdi]{Maria De Iorio}
\author[aj]{Ajay Jasra}
\author[mch]{Marc Chadeau-Hyam}

\address[gc,mch]{Department of Epidemiology and Biostatistics, Imperial College London\\
                 London W2 1PG, UK}
\address[mdi]{Department of Statistical Science, University College London\\
              London WC1E 6BT, UK}
\address[aj]{Department of Statistics and Applied Probability, National University of Singapore\\
             Singapore 117546, SG}

\begin{abstract}
The simulation of genealogical trees backwards in time, from observations up to
the most recent common ancestor (MRCA), is hindered by the fact that, while
approaching the root of the tree, coalescent events become rarer, with a
corresponding increase in computation time.
The recently proposed ``Time Machine'' tackles this issue by stopping the
simulation of the tree before reaching the MRCA and correcting for the induced
bias.
We present a computationally efficient implementation of this approach that
exploits multithreading.
\end{abstract}

\begin{keyword}
population genetics \sep importance sampling \sep algorithms \sep parallel computing
\end{keyword}

\end{frontmatter}

\section{Introduction}
There is currently much interest in performing ancestral inference from
molecular data: experimental advances have led to new types and greater
availability of data, giving rise to new challenges for the mathematical
community.
Advanced stochastic models have been developed to capture the complexity in this
data, in turn leading to further challenges to achieve computational efficiency.

Many of these models are formulated as hidden (or unobserved) Markov chains, and
the likelihood of the data is obtained by taking the expectation over all
possible realizations of the underlying Markov process.
An important example is the coalescent model: this hidden Markov chain is
typically stopped at a random time, which usually corresponds to the time to
reach the most recent common ancestor (MRCA).
Computing the marginal likelihood associated with the observed data is, in
general, not analytically tractable.
Consequently, estimating unknown parameters by maximum likelihood requires a
numerical approximation scheme, which is most conventionally based upon
importance sampling as in \cite{stephens_jrss_2000}.
While tracking the tree backwards in time, the population size diminishes,
leading to fewer coalescent events and thus to an increased stopping time.
Furthermore, the information relevant to statistical inference also diminishes,
resulting in importance weights with substantial variance.
For these reasons, \citet{jasra_prsa_2011} proposed to stop the simulation
before reaching the MRCA, and to account for the resulting bias in the
likelihood estimates.
This method brings about substantial reductions in both computation time and
variance of the estimates.

The \texttt{TimeMachine} package is a computationally efficient implementation
of the biased numerical likelihood approximation of \citet{jasra_prsa_2011}.
It provides estimates of the likelihood surface, and can additionally estimate
the mutation rate $\mu$ by likelihood maximisation.

\section{Algorithm}
The algorithm simulates genealogical trees backwards in time, from a given
initial population up to the point when there are $N\!\geq\!1$ sequences left.
The likelihood is estimated by averaging over many independent simulations.
The special case $N\!=\!1$ corresponds to the approach of
\citet{stephens_jrss_2000}, whereas $N\!>\!1$ corresponds to the
``Time Machine'' of \citet{jasra_prsa_2011}.

The main steps of the iterative algorithm are described below; detailed
calculations can be found in the software documentation.
\begin{enumerate}
    \item Sample the offspring type $i$ with probability proportional to the
          number of individuals of that type in the population;
    \item Sample the ancestor type $j$: an offspring of type $i$ might have
          arisen from an ancestor of type $j$ through either a coalescent event
          or a $j \to i$ mutation (with $j$ possibly equal to $i$);
    \item Update the population size within each type;
    \item Compute the contribution to the likelihood of the simulated event and
          update the likelihood;
    \item Assess the stopping criterion: if $N\!>\!1$, stop if the desired
          population size has been reached; otherwise, repeat the above steps
          until only two individuals are left in the population, at which point
          mutations are exclusively simulated until both remaining sequences are
          of the same type;
    \item If $N\!>\!1$, correct the likelihood to account for the bias induced
          by stopping the simulation before reaching the MRCA.
\end{enumerate}

\section{Implementation}
The software is released as an \texttt{R} package, and is freely available on
the Comprehensive \texttt{R} Archive Network (CRAN) package repository.

Most functions are internally implemented in C for optimal performance.
Thanks to the OpenMP\textsuperscript{\textregistered} API, simulations can be
run in parallel threads on machines with multiple cores or processors.
To ensure independence, each thread uses a different seed to initialise its copy
of the pseudorandom number generator of \cite{panneton_atms_2006}.

\section{Features}
The \texttt{TimeMachine} can be used on:
\begin{inparaenum}[(a)]
    \item a user-specified initial population;
    \item a population sampled from a multinomial model with probability vector
          corresponding to the stationary distribution associated with a given
          unit transition matrix.
\end{inparaenum}

The main output returned by the \texttt{TimeMachine} is a list containing the
(possibly corrected) log-likelihoods and corresponding correction terms, as well
as the per-simulation and total computation times.
To allow the reconstruction of each simulated tree, the sequential event number
(SEN) of each coalescent event and the distribution of types at the last
iteration are also stored.
An extensive description of all available outputs is given in the documentation.

The \texttt{TimeMachine} package also includes a maximum-likelihood estimation
procedure for the mutation rate $\mu$ based on the \texttt{R} built-in
\texttt{optimize} function.

\begin{figure*}[htb]
    \centering
    \includegraphics[width=\textwidth]{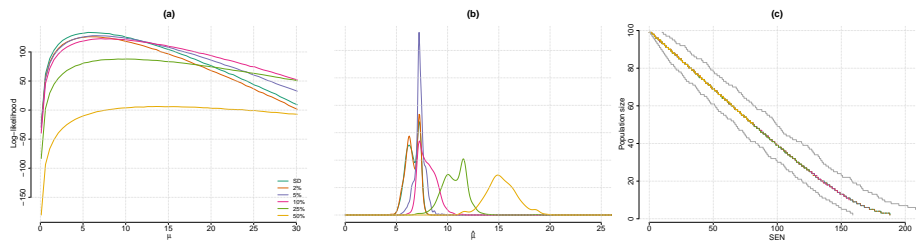}
    \caption{
        Outputs based on populations of 100 individuals sampled under a PDM and
        six values of $N$:
        (a) estimated log-likelihood surfaces for a fixed population and
            different values of $\mu$;
        (b) density of $\hat{\mu}$ over 1,000 populations;
        (c) median population size as a function of SEN (grey lines correspond
            to extrema for $N\!=\!1$).
    }
    \label{fig:outputs}
\end{figure*}

\section{Application}
We illustrate the use of the \texttt{TimeMachine} package on a set of 1,000
initial populations, each comprising 100 individuals divided into
$2^{10} = 1024$ types, sampled under the parent-dependent model (PDM) already
introduced in \citet{jasra_prsa_2011}.
We investigated several scenarios corresponding to $N\!=\!1, 2, 5, 10, 25, 50$.

In Fig.~\ref{fig:outputs}a we plot the estimated log-likelihoods (averaged over
10,000 independent simulations) for a fixed population and 60 values of $\mu$,
ranging from 0.1 to 30.1.
As shown, the log-likelihood surfaces for $N\!\leq\!10$ closely match the exact
case $N\!=\!1$, suggesting that stopping simulations beforehand only marginally
affects the precision and reliability of the estimates, while bringing about
significant improvements in computation time (two-fold reduction in computation
time for $N\!=\!10$).

Fig.~\ref{fig:outputs}b shows the density of the maximum-likelihood estimates of
the mutation rate $\hat{\mu}$ obtained for the 1,000 initial populations and
different values of $N$ (as above).
Distributions are narrow, with a dispersion that increases with $N$.
Moreover, they share the same mode for $N\!\leq\!10$, again suggesting that the
accuracy of the estimation is only marginally affected in these cases.

Finally, in Fig.~\ref{fig:outputs}c we summarise the history of each simulation
for a fixed population by plotting the median population size as a function of
the sequential event number (SEN).
As expected, trajectories are consistently simulated across different values of
$N$, with a plateau towards bigger SEN that is associated to rarer coalescent
events and greater variation.

\section{Conclusion}
We have presented a computationally efficient \texttt{R} implementation of the
``Time Machine'' recently proposed by \citet{jasra_prsa_2011}.
Given the ability to remove the random time for the hidden Markov chain, this
implementation is amenable to practical testing of the bias that was
theoretically and empirically investigated in \citet{jasra_prsa_2011}.

The main limiting factor of the software is the memory required to store the
transition matrix between types, which is of size $2^{L} \times 2^{L}$ for $L$
loci.
This restricts usage on modern personal computers to approximately 15 loci.
This issue could be addressed here by only considering a subset of biologically
meaningful types.
The current implementation could incorporate this change at the cost of minimal
edits, and thus represents a first step towards the development of a general
framework for approximate inference in population genetics based on the
stopping principle.
It is our hope that our contribution will allow biologists and experimental
scientists to analyse their data more efficiently.

\bibliographystyle{plainnat}
\bibliography{time_machine}

\end{document}